\documentclass[aps,prd,groupedaddress,superscriptaddress,11pt,nofootinbib,preprintnumbers]{revtex4-2}

\usepackage{dsfont}
\usepackage{natbib}
\usepackage{url}
\usepackage{color}
\usepackage{fullpage}
\usepackage[top=2cm, bottom=4.5cm, left=2.5cm, right=2.5cm]{geometry}
\usepackage{amsmath,amsthm,amsfonts,amssymb,amscd}
\usepackage{lastpage}
\usepackage{enumerate}
\usepackage[shortlabels]{enumitem}
\usepackage{fancyhdr}
\usepackage{mathrsfs}
\usepackage{xcolor}
\usepackage{graphicx}
\usepackage{listings}
\usepackage{hyperref}

\hypersetup{%
  colorlinks=true,
  linkcolor=blue,
  linkbordercolor={0 0 1}
}

\newcommand{\be}{\begin{equation}}
\newcommand{\ee}{\end{equation}}
\newcommand{\bea}{\begin{eqnarray}}
\newcommand{\eea}{\end{eqnarray}}
\newcommand{\nn}{\nonumber}

\newcommand{\pp}{P_{\phi}}
\newcommand{\pv}{P_{\varphi}}
\newcommand{\ha}{\mathcal{H}}
\newcommand{\pt}{\widetilde{\phi}}
\newcommand{\ppt}{\widetilde{P}_{\phi}}
\newcommand{\TT}{\widetilde{T}}
\newcommand{\PT}{\widetilde{P}_{T}}
\renewcommand{\H}{\mathcal{H}}
\newcommand{\sq}{\sqrt{q}}

\begin{document}

\title{From global time to local physics}

\bigskip

\author{Syed Moeez Hassan}
\email{syed\_hassan@lums.edu.pk}
\affiliation{Department of Physics, Syed Babar Ali School of Science and Engineering, Lahore University of Management Sciences, Lahore 54792, Pakistan} 

\author{Viqar Husain}
\email{vhusain@unb.ca}
\affiliation{Department of Mathematics and Statistics, University of New Brunswick, Fredericton, NB, Canada E3B 5A3}

\author{Babar Qureshi}
\email{bqureshi@iba.edu.pk}
\affiliation{Department of Mathematical Sciences, IBA, Karachi, Pakistan}

\date{\today}

\begin{abstract}

\vskip 1cm

Global time is a gauge or relational choice of time variable in canonical gravity. Local time is the time used in a flat patch of spacetime. We compare the dynamics of a scalar field with respect to choices of global time and Minkowski patch time in an expanding cosmology. Our main results are that evolutions starting from the same initial conditions are similar on the time scales of terrestrial experiments, and that global time leads to a mechanism for evolving coupling constants.

\end{abstract}

\maketitle
\vskip 0.5cm
\section{Introduction}

Clocks are oscillatory physical systems. For all terrestrial experiments their readings coincide with either Newtonian absolute time or the special relativistic time of a local inertial frame. For cosmological theories and observations described by a spacetime metric theory, the notion of time is neither of these. Instead, it is a ``time" constructed out of  dynamical variables, such as the Hubble time $T_H = a(t)/\dot{a}(t)$ coming from the scale factor $a(t)$ of homogeneous isotropic models. This is fundamentally different from the notion of time used in laboratory experiments, where the clock is ``external" to the system under observation, not interacting with the system, and monotonically increasing on the scale of laboratory experiments.  

These differences between local clocks external to the system used in labs and global internal clocks such as Hubble time, pose interesting foundational questions about the origin of time, in particular the compatibility of global versus local notions of time. Every laboratory system is a flat Minkowski patch of an expanding Universe in which the time scale of experiments is a tiny fraction of the age of the Universe as measured by Hubble time. 

From the perspective of general relativity coupled to matter described as a phase space system, there is no natural notion of time that might describe physical evolution. It is well-known that this ``problem of time" \cite{Isham:1992ms,Kuchar:1991qf,Anderson:2010xm,Anderson:2017jij} arises ultimately from the fact that the Hamiltonian of the system is a constraint condition on the phase space -- it represents a fine balance between the matter and geometry variables. The problem of time  is ``resolved" through a choice of time and other coordinate gauges, or by introducing a mechanism of relational time where one phase variable is used as the clock to measure evolution of other phase space variables. We refer to such a choice as ``global time". 

In contrast, the time variable that appears in classical and quantum physics is either Newtonian absolute time or the time in a special relativistic inertial frame. It happens to coincide with the time we measure despite there being no physical description of the clock itself in these theories.  This comes from a fixed background metric structure and coincides with the natural time in a Minkowski patch in general relativity. We refer to this as ``local time".  Related questions concerning what is a physical clock and what is a cosmological one have been discussed in the philosophy of science literature, e.g.  \cite{pittphilsci22565,Rugh:2008xs,Rugh:2016olq}.

In this paper we show that, starting from general relativity coupled to matter, there is a way to recover the dynamics with respect to Minkowski Killing-time  only for certain  choices of global clocks. We also show that a global choice of clock also provides a natural mechanism for ``evolving" coupling constants; this is unlike other ideas and mechanisms for evolving constants \cite{Dirac:1937ti,Bekenstein:1982eu,Sandvik:2001rv,Barrow:2005hw, Uzan2011}. All our results are obtained  for general relativity coupled to dust and a scalar field; while our results are classical, this is a well studied model at the quantum level \cite{Husain:2019nym,Gielen:2021igw}. 

An interesting related question is whether a global choice of time, either fixed through matter or geometric variables, can be consistent with Lorentz invariance. This question has been posed for general relativity coupled to a scalar field where the latter is chosen as the clock \cite{Smolin:1993ka}; this choice gives a corresponding physical Hamiltonian that leads in a certain expansion to a theory that is ``Poincare invariant, even if the starting point is not," as the author notes. We also address this question, and show that the global time defined by dust respects Poincare invariance, without any approximation. 

In the next section (\ref{sec-gr}) we review the canonical formulation of general relativity minimally coupled to a pressureless dust and scalar field. In Sec. \ref{sec-time} we consider this model in the volume and dust time gauges in the cosmological setting and compare the evolution of the scalar field with these two clocks with the same evolution in  Minkowski spacetime. In Sec. \ref{sec-beyond-cosmo}, we discuss whether such a procedure is possible beyond cosmological models in the full theory. We summarize and discuss our conclusions in the last section (\ref{sec-disc}).

\section{Scalar field and dust in general relativity} \label{sec-gr}

We start with general relativity written in the Arnowitt-Deser-Misner (ADM) form minimally coupled to a pressureless dust field $T$, and a scalar field $\phi$ with a potential $V(\phi)$ (we work in $c = \hbar = 8 \pi G = 1$ units),
\be
\label{full-theory}
S = \int d^3x dt ~ [ \pi^{ab} \dot{q}_{ab} + \pp \dot{\phi} + P_T \dot{T} - N(\H_G + \H_{\phi} + \H_D) - N^a (C^G_a + C^{\phi}_a + C^D_a)],
\ee
where,
\bea
\label{full-ham}
\H_G &=& \dfrac{1}{\sq} \Bigg(\pi^{ab} \pi_{ab} - \frac{1}{2} \pi^2 \Bigg) + \sq ( \Lambda - R), \nn \\
\H_{\phi} &=& \dfrac{\pp^2}{2 \sq} + \dfrac{1}{2} \sq q^{ab} \partial_a \phi \partial_b \phi + \sq V(\phi), \nn \\
\H_D &=& P_T \sqrt{1 + q^{ab} \partial_a T \partial_b T }, \\
C^{G}_a &=& D_b \pi^b_a, \nn \\
C^{\phi}_a &=& \pp \partial_a \phi, \nn \\
C_a^D &=& -P_T \partial_a T,
\eea
$(q_{ab}, \pi^{ab})$, $(\phi, \pp)$ and $(T, P_T)$ are the gravitational, scalar field and dust phase space variables respectively, $q$ is the determinant of the spatial metric $q_{ab}$, $\pi$ is the trace of $\pi^{ab}$, $R$ is the 3-Ricci scalar curvature, $\Lambda$ is the cosmological constant, $N$ is the lapse and $N^a$ is the shift.

We now reduce to a spatially flat ($k=0$) Friedmann-Lemaitre-Robertson-Walker (FLRW) Universe with the ansatz
\be
q_{ab} = a^2(t) e_{ab} , ~ \pi^{ab} = \dfrac{P_a(t)}{6a(t)} e^{ab},
\ee
where $a(t)$ is the scale factor, $P_a$ is the momentum canonically conjugate to the scale factor, and $e_{ab} = \text{diag}(1,1,1)$ is the flat metric. The scalar and dust fields must therefore also be homogeneous:
\be
\phi = \phi(t) ~ , ~ \pp = \pp(t),
\ee
\be
T = T(t), ~ P_T = P_T(t).
\ee
With this homogeneity ansatz, the diffeomorphism constraint vanishes identically and we are left with
\be
S = V_0 \int dt ~ \Bigg( P_a \dot{a} + \pp \dot{\phi} + P_T \dot{T} - NH \Bigg),
\ee
after the spatial integration, with the Hamiltonian constraint
\be
\label{Ham-constr-cosm}
H =  -\dfrac{P_a^2}{24 a}  + a^3 \Lambda  + \dfrac{\pp^2}{2 a^3} + a^3 V(\phi) + P_T \approx 0,
\ee
and $ V_0 = \displaystyle \int ~ d^3x $.

The FLRW line element is invariant under the scaling 
\be
x\rightarrow s x,\quad a\rightarrow a/s. 
\ee
This induces the following scale transformations of variables in  the canonical action 
\be
\left( a, P_a, \phi, P_\phi, T, P_T \right) \rightarrow \left( a/s, P_a/s^2, \phi, P_\phi/s^3, T, P_T/s^3 \right) 
\ee
The scalar potential $V(\phi)$ and the cosmological constant $\Lambda$ are invariant under this scale transformation. This naturally suggests the following definitions of scale-invariant variables:
\be
a = V_0^{1/3} a, ~~ P_a = V_0^{2/3} P_a, ~~ \pp = V_0 \pp, ~~ P_T = V_0 P_T
\ee
(Effectively, this sets $V_0 = 1$ in the action, with everything taking the same form as before but now in terms of scale-invariant variables).

\section{Local physics from global time} \label{sec-time}

A global time gauge fixing proceeds by choosing some combination of the phase space variables as a clock, i.e. $f(p,q) =t$. Then the Hamiltonian constraint is solved (strongly) for the variable canonically conjugate to this choice of time. This leads to a non-vanishing physical Hamiltonian from which dynamics can be derived. We illustrate this procedure in the next subsection for two choices of time.

\subsection{Volume time}

The first global clock we consider is made up of the gravitational configuration degree of freedom $t=f(a)$. For now, we leave this function arbitrary. This variable is canonically conjugate to  $P_a/f'(a)$ (where the prime indicates a derivative with respect to $a$), since their canonical Poisson bracket is $\Big\{f(a), P_a/f'(a) \Big\} = 1$. The physical Hamiltonian is the negative of the variable canonically conjugate to time, and is  obtained after solving the Hamiltonian constraint and eliminating any gravitational degrees of freedom through $a = f^{-1}(t)$,
\be
H_p = -\dfrac{P_a}{f'(a)} \bigg\rvert_{t = f(a), \ha = 0}.
\ee

The lapse function $N$ is fixed by the requirement that the time gauge fixing must be dynamically preserved,
\be
1 = \dot{t} = \{t, NH \} = N \{t, H \}
\ee
and the gravitational symplectic term becomes
\be
P_a \dot{a} = P_a \dfrac{\dot{t}}{f'(a)} = -H_p.
\ee
These steps lead to the (time) gauge fixed action,
\be
S^{GF} = \int dt \Big( \pp \dot{\phi} + P_T \dot{T} - H_p \Big).
\ee
Solving for $H_p$ from the Hamiltonian constraint gives
\be
H_p = \sqrt{24} ~ \dfrac{a^2}{f'(a)} ~ \sqrt{ \Lambda + \Bigg[ \dfrac{\pp^2}{2 a^6} + V(\phi) + \dfrac{P_T}{a^3} \Bigg] } \Bigg\rvert_{t = f(a)} \equiv \sqrt{24} ~ \dfrac{a^2}{f'(a)} ~ \sqrt{\rho_{\Lambda} + \rho_M} \Bigg\rvert_{t = f(a)},
\ee
where in the last line, we have identified the usual matter and cosmological constant energy densities. This completes the procedure of fixing a global time gauge and obtaining the corresponding physical Hamiltonian. We note two key features of this Hamiltonian: it is a square root, and it is explicitly time dependent (through substituting $a = f^{-1}(t)$).

We now come to the central assumption in this subsection that will allow us to make a connection with local physics. In its current form (with the square root), the Hamiltonian does not look like any known Hamiltonian of a theory describing local physics. To obtain a standard form -- a sum of kinetic and potential terms -- we expand the square root  assuming that 
\be
\label{exp-assum}
\rho_M ~<~ \rho_{\Lambda}.
\ee

This gives, 
\be
H_p = \Bigg( 2 \sqrt{6 \Lambda} \dfrac{a^2}{f'(a)}\Bigg) + \sqrt{\dfrac{6}{\Lambda}} \dfrac{a^2}{f'(a)} \Bigg[ \dfrac{\pp^2}{2 a^6} + V(\phi) + \dfrac{P_T}{a^3} \Bigg].
\ee

The first term in the parenthesis above, is purely a function of time, and will lead to a boundary term in the action, which can be ignored; the rest takes the form,
\be
H_p = \dfrac{\pp^2}{2} \left[\dfrac{1}{x^2(t)} \right] + V(\phi) \left[y^2(t) \right] + P_T \left[ \dfrac{1}{z(t)} \right],
\ee
where 
\bea 
x^2(t) &=& \sqrt{\frac{\Lambda}{6}} a^4 f'(a)\nn\\
y(t) &=& x(t) \left( \sqrt{\frac{6}{\Lambda}} \frac{1}{a f'(a)} \right)\nn\\
z(t) &=& a f'(a) \sqrt{\frac{\Lambda}{6}}.
\eea
For concreteness, we fix the potential to be  $V(\phi) = \frac{1}{2} m^2 \phi^2 + \lambda \phi^4$.

For the purpose of comparing the dynamics of the scalar field with the standard one on flat spacetime, we seek a time-dependent canonical transformation. This is accomplished by defining
\be
\ppt = \dfrac{\pp}{x(t)}, ~~ \pt = \phi x(t); ~~ \PT = \dfrac{P_T}{z(t)}, ~~ \TT = T z(t),
\ee
which is generated by the function $F_2(\phi, \ppt, t) = \phi \ppt x(t)$ for the scalar field (and similarly for dust: $F_2(T, \PT, t) = T \PT z(t)$). Under this transformation, the action becomes,
\be
S = \int dt ~ [\ppt \dot{\pt} + \PT \dot{\TT} - \widetilde{H}_p]
\ee
with,
\be
\widetilde{H}_p = \dfrac{\ppt^2}{2} + \dfrac{1}{2} \widetilde{m}^2 \pt^2 + \widetilde{\lambda} \pt^4 + \PT  + \pt \ppt ~ \Big(\dfrac{\dot{x}}{x} \Big) + \TT \PT ~ \Big(\dfrac{\dot{z}}{z} \Big)
\ee
where,
\be
\widetilde{m} = m \left[\sqrt{\dfrac{6}{\Lambda}} \dfrac{1}{a f'(a)} \right], ~~ \widetilde{\lambda} = \lambda \left[\sqrt{\dfrac{6}{\Lambda}} \dfrac{1}{a^2 f'(a)} \right]^3, ~~ \left(\dfrac{\dot{x}}{x} \right) = \dot{a} \left(\dfrac{2}{a} + \dfrac{f''(a)}{2f'(a)} \right), ~~ \left(\dfrac{\dot{z}}{z} \right) = \dot{a} \left(\dfrac{1}{a} + \dfrac{f''(a)}{f'(a)} \right)
\ee

We now specialize to a particular choice of time, the volume time, $t=f(a)=a^3$, which measures time as the size (volume) of the universe. This gives
\be
\label{vol-ham}
\widetilde{H}_p = \dfrac{\ppt^2}{2} + \dfrac{1}{2} \widetilde{m}^2(t) \pt^2 + \widetilde{\lambda}(t) \pt^4 + \PT  + \dfrac{\pt \ppt}{t} + \dfrac{\TT \PT}{t},
\ee
where
\be
\label{time-dep-const}
\widetilde{m}(t) = \dfrac{m}{t} \sqrt{\dfrac{2}{3\Lambda}}, ~ \widetilde{\lambda}(t) = \dfrac{\lambda}{t^4} \left(\sqrt{\dfrac{2}{3\Lambda}}\right)^{3}.
\ee
This Hamiltonian has some interesting features: (i) the coupling constants (mass $m$, and $\lambda$) acquire a time dependence; (ii) if the universe is expanding, $t$ increases monotonically, which means that these coupling constants are \textit{dynamically} driven to zero; (iii) apart from the last two terms (which also have a $1/t$ dependence), this Hamiltonian is just the usual Hamiltonian of a scalar field and a dust field (albeit with time dependent coupling constants) that one would write down in flat spacetime.

It is this last point that we wish to emphasize further here: starting from a global choice of time (volume time), we have been able to get to a theory describing physics with respect to Minkowski patch time. To see how closely this reduced theory matches local physics, we look at the dynamics through the equations of motion (an over-dot indicates a derivative with respect to the global time $t$),
\bea
\dot{\pt} &=& \ppt + \dfrac{\pt}{t} \label{scalareq1} \\
\dot{\ppt} &=& -\widetilde{m}^2(t) \pt -4 \widetilde{\lambda}(t) \pt^3 - \dfrac{\ppt}{t} \label{scalareq2} \\
\dot{\TT} &=& 1 + \dfrac{\TT}{t} \\
\dot{\PT} &=& - \dfrac{\PT}{t}.
\eea
The Hamiltonian also evolves in time: $\dot{\widetilde{H}_p} = \dfrac{\pt}{t} \dot{\ppt} + \dfrac{\TT}{t} \dot{\PT}$. We want to compare this dynamics to the dynamics of a theory of a scalar field and a dust field written down in a local region of spacetime (laboratory) with respect to the (local) Minkowski patch time. It is clear from the above equations that (apart from the time-dependence of the coupling constants $m$ and $\lambda$) the last term in each of them (with the $1/t$ dependence) is responsible for any deviations from what might be seen locally. However, all of these terms, and even the coupling constants, contain a $1/t$ dependence or higher. This means that as the universe expands, $t$ (which is the volume of the universe) increases, and all of these terms go to zero, making any deviations from local physics also go to zero.

To elucidate the role of this last term, let us look at the solutions to these equations. There is no coupling between the scalar and the dust fields, therefore, we can solve them separately.  The dust equations have solutions,
\bea
\TT &=& t (\log(t) + c_1), \\
\PT &=& \dfrac{c_2}{t},
\eea
where $c_1$ and $c_2$ are some constants of integration. To connect this with local physics, the global volume time $t$ can be written as $t = t_0 + \delta t$ where $t_0$ is the time now, and $\delta t$ is the measure of time in some small neighbourhood of $t_0$. The key point is that in this global volume time gauge, $t_0$ is very large (the universe is very large) and $\delta t$ is very small (the size of the universe does not change by much over the timescales of local experiments). For instance, if we plug in the current numerical values, we find (in natural units), $t_0 \sim 10^{184}$. If we consider local experiments on timescales of $\sim 1000$ years, we have $\delta t \sim 10^{54}$, which gives $\delta t/t_0 \sim 10^{-130}$. This allows us to expand the above solutions near $t=t_0$ to obtain,
\bea
\TT &\approx& [(c_1+\log(t_0))t_0] + [(c_1+\log(t_0)+1)t_0] \left( \dfrac{\delta t}{t_0} \right) + \mathcal{O}\left( \left( \dfrac{\delta t}{t_0} \right)^2 \right) \\
&\approx& c_1^{'} + \tilde{c}_1 \delta t,\\
\PT &\approx& \left[ \dfrac{c_2}{t_0} \right] - \left[ \dfrac{c_2}{t_0} \right] \left( \dfrac{\delta t}{t_0} \right) + \mathcal{O}\left( \left( \dfrac{\delta t}{t_0} \right)^2 \right) \\
&\approx& c_2^{'}
\eea
where in the last equation for $\PT$, we have ignored the linear term because it contains a factor of $1/t_0^2$ in the denominator. Such behaviour is not present in the linear term in $\TT$, hence we must keep that term. These results reproduce exactly the behaviour we expect from a free dust field in flat spacetime.

\begin{figure*}
\includegraphics[width=0.6\textwidth]{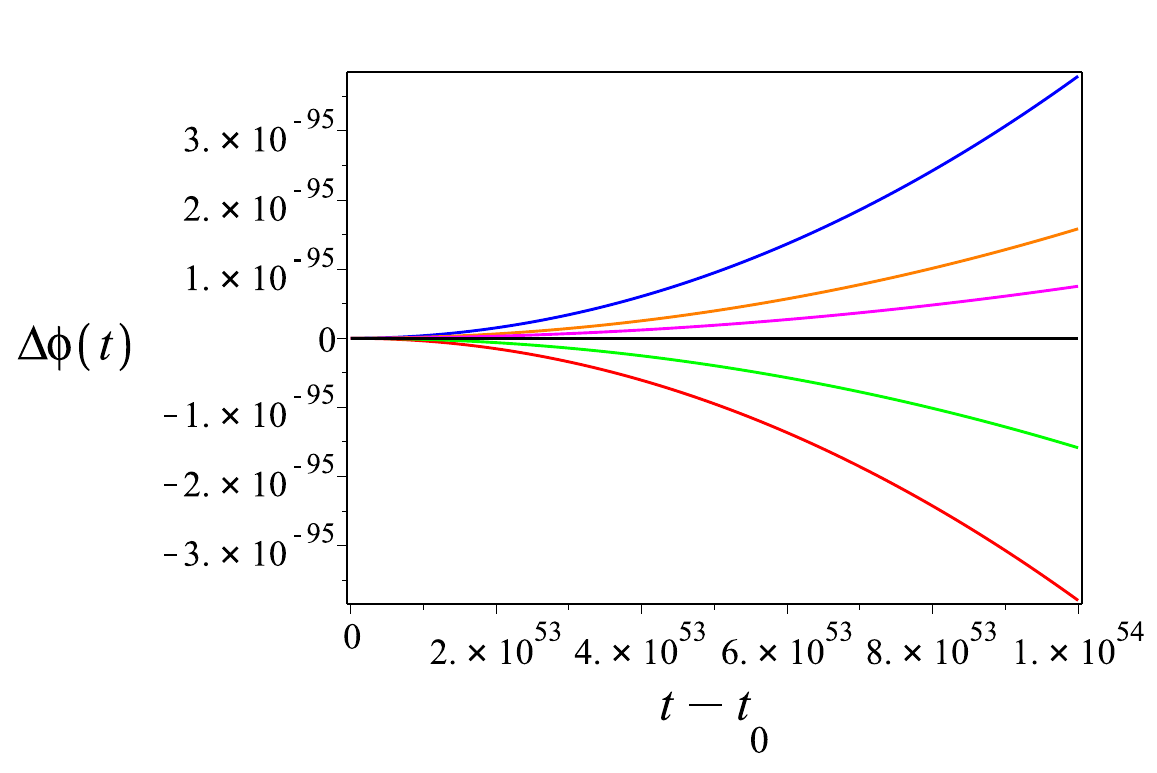}
\caption{\label{fig-phidiff} The time evolution of $\Delta \phi = \pt - \phi_{loc}$. Different colours represent different initial conditions.}
\end{figure*}

\begin{figure*}
\includegraphics[width=0.6\textwidth]{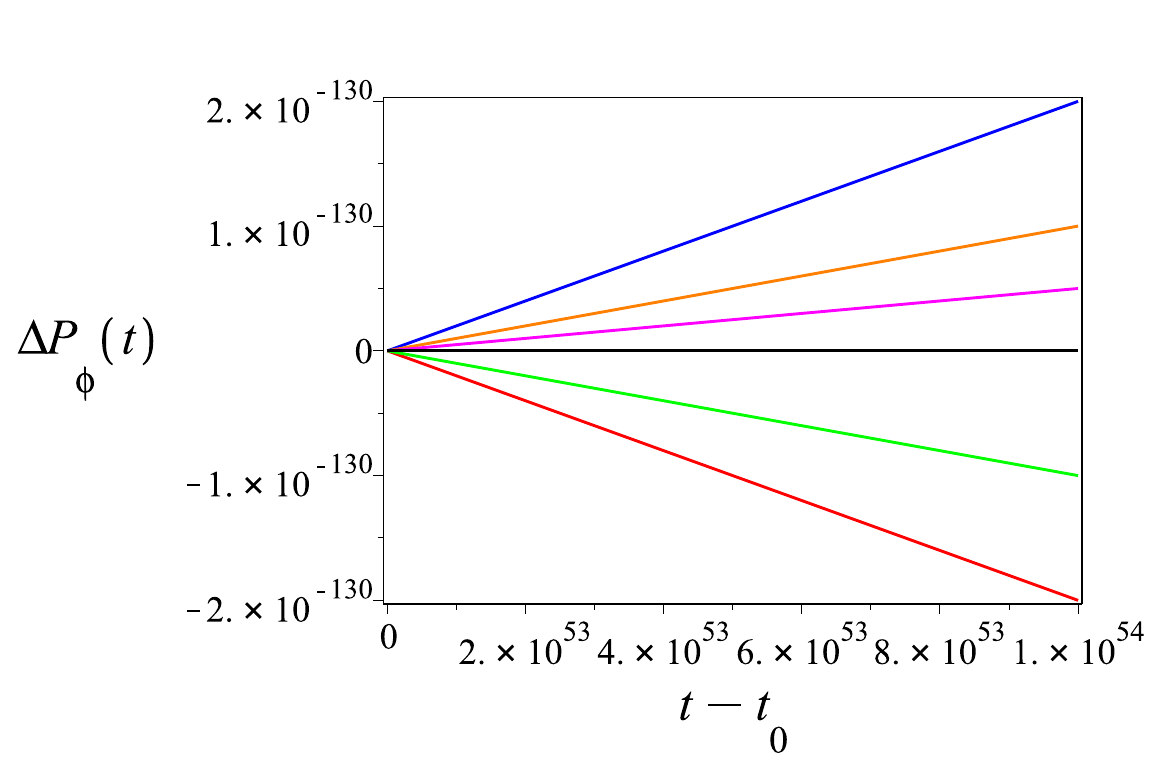}
\caption{\label{fig-pdiff} The time evolution of $\Delta P_{\phi} = \ppt - P_{loc}$. Different colours represent different initial conditions.}
\end{figure*}

\begin{figure*}
\includegraphics[width=0.6\textwidth]{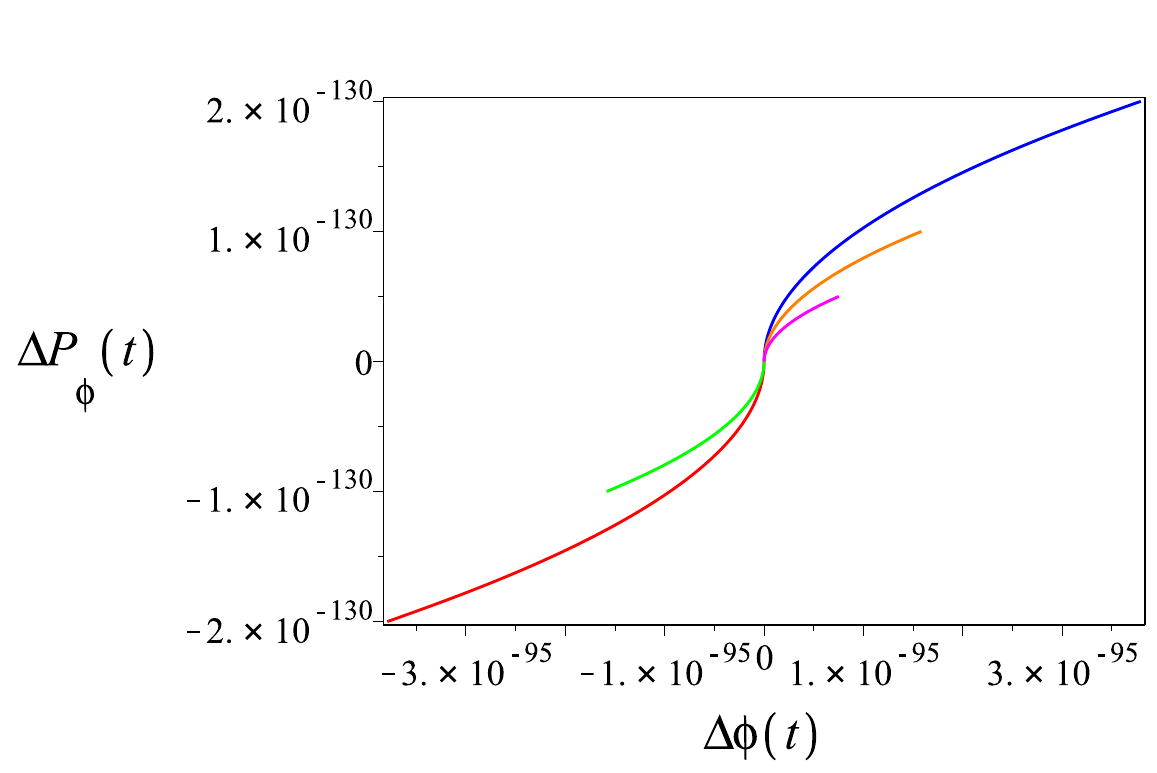}
\caption{\label{fig-phasediff} Phase portrait of $(\Delta \phi, \Delta P_{\phi})$. Different colours represent different initial conditions.}
\end{figure*}

The analysis for the scalar field proceeds in a similar fashion. If there is no potential ($m=\lambda=0$), then we can find exact analytical solutions and expand them near $t = t_0$ to recover flat space behavior. In the presence of a potential, we note that both the mass and the quartic coupling terms become very small when the universe is large due to their time dependence (Eqn. \ref{time-dep-const}). To look at the exact behaviour, we numerically solved the scalar field equations of motion (Eqns. \ref{scalareq1}, \ref{scalareq2}), and compared these solutions with the usual dynamics of a (local) scalar field in Minkowski spacetime ($\phi_{loc},P_{loc}$) with mass $m_{loc}$ and quartic coupling $\lambda_{loc}$. These new parameters ($m_{loc}$ and $\lambda_{loc}$) were chosen to match the values the time dependent coupling constants $m(t)$ and $\lambda(t)$ (in Eqn. \ref{time-dep-const}) would take now (at $t=t_0$). Our results are shown in Figures (\ref{fig-phidiff} - \ref{fig-phasediff}).

Figure \ref{fig-phidiff} shows the time evolution of $\Delta \phi = \pt - \phi_{loc}$ (where $\pt$ is the global scalar field) on timescales of local experiments. Similarly, Figure \ref{fig-pdiff} shows the time evolution of $\Delta P_{\phi} = \ppt - P_{loc}$ on these timescales. As is clear from these figures, there is negligible difference between the dynamics of the global scalar field and the local one during this period of time.

Figure \ref{fig-phasediff} shows the differences between the two fields on a phase portrait with a wide range of initial conditions. The phase portrait is plotted with the differences ($\Delta \phi$ and $\Delta P_{\phi}$) to show any deviations. Again, there is almost no difference on these scales, and the results are robust with respect to choices of different initial conditions.

\begin{figure*}
\includegraphics[width=0.6\textwidth]{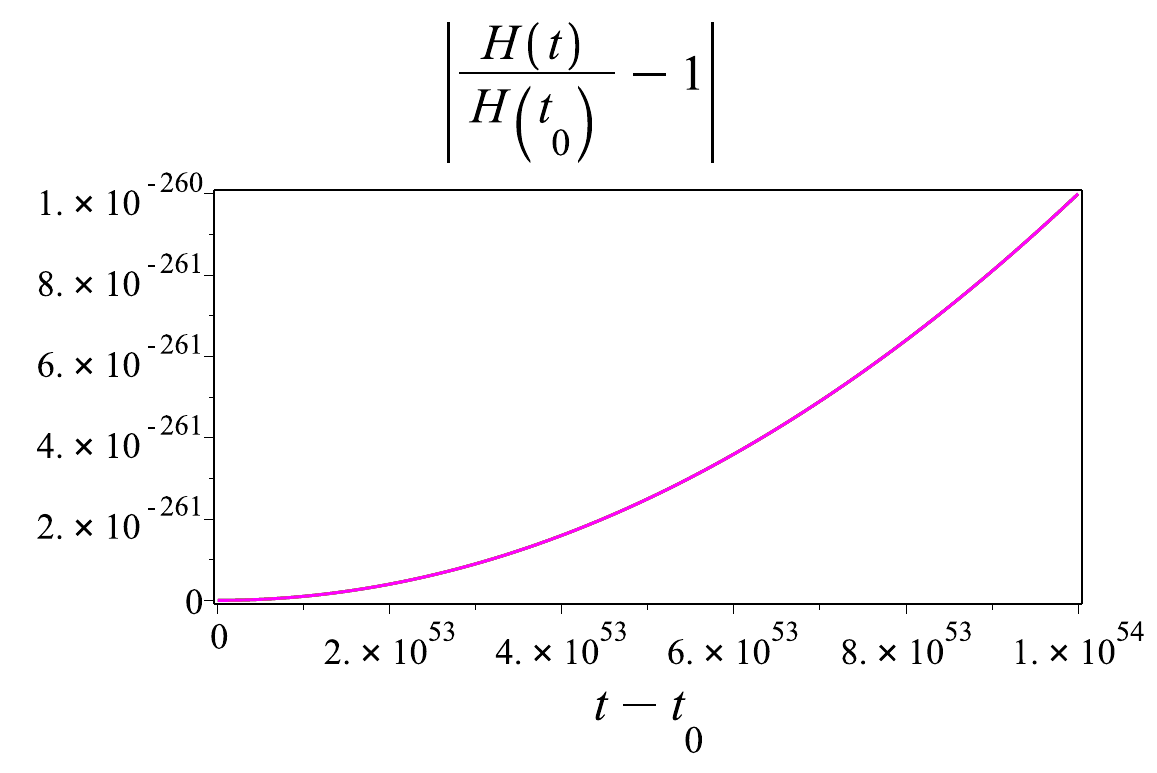}
\caption{\label{fig-ham} Time evolution of the Hamiltonian. The evolution with different initial conditions is very close to one another for any differences to be visible on this scale.}
\end{figure*}

Since the Hamiltonian obtained from this procedure (Eqn. \ref{vol-ham}) is explicitly time dependent, whereas the Hamiltonian describing the dynamics of a local scalar field is not, we also computed the time evolution of this Hamiltonian on relevant timescales. The result is shown in Figure \ref{fig-ham} (different colors represent different initial conditions) which shows that although there is time evolution in the Hamiltonian, it is negligible on these timescales, ensuring that for all practical purposes, it does behave as the usual Hamiltonian of a scalar field in flat spacetime.

\begin{figure*}
\includegraphics[width=0.6\textwidth]{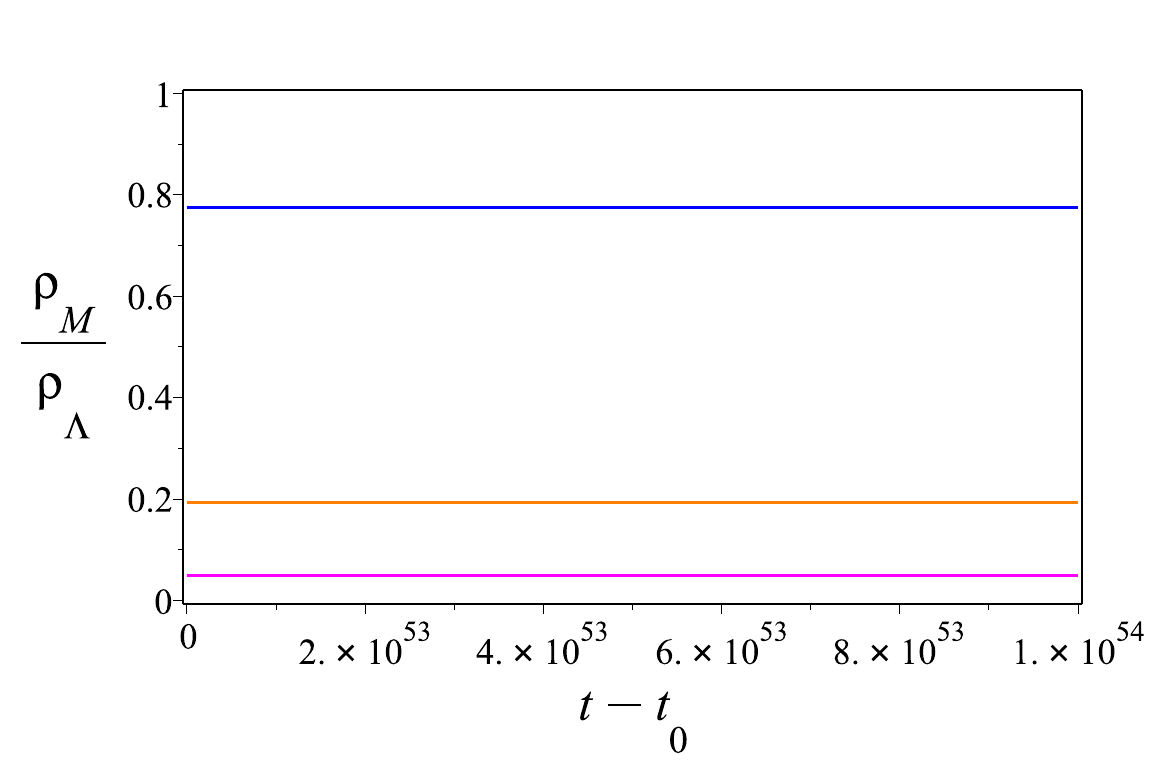}
\caption{\label{fig-exp} Time evolution of $\rho_M/\rho_{\Lambda}$. It should remain below 1 to justify our assumption of expanding the square root. Different colours represent different initial conditions, but some are too close to one another to be visible on this scale.}
\end{figure*}

Finally, we also test the validity of our initial assumption for the square root expansion (Eqn. \ref{exp-assum}), and compute the evolution of $\rho_M/\rho_{\Lambda}$ on relevant timescales for various initial conditions. Figure \ref{fig-exp} shows the results which indicate that it is indeed a good assumption for a large range of initial conditions. Hence, we have shown that using the volume of the universe as global time, we can recover local dynamics of a scalar field. In the next subsection, we show how this can also be achieved with a special matter clock: dust as time.

\subsection{Dust time}

We now choose a matter field (dust) as our clock: $t=T$ \cite{Husain:2011tk}. The momentum conjugate to this is $P_T$, and therefore, the Physical Hamiltonian is,
\be
H_p = -\dfrac{P_a^2}{24 a} + a^3 \Lambda + \dfrac{\pp^2}{2 a^3} + a^3 V(\phi),
\ee
with the (time) gauge-fixed action,
\be
S^{GF} = \int dt [P_a \dot{a} + \pp \dot{\phi} - H_p].
\ee
Notice that this physical Hamiltonian is not a square-root as the Hamiltonian constraint (Eqn. \ref{Ham-constr-cosm}) is linear in the dust momentum. With this choice  the scalar field ($\phi$) and gravity ($a$) are dynamical.

To connect with local physics (where gravity is not dynamical), we consider a fixed background limit of this theory (this is equivalent to doing quantum field theory on a fixed background but in the dust time gauge). Since we are fixing the background, we can evaluate the first (symplectic) term in the action ($P_a \dot{a}$) on this fixed background, and include it in our physical Hamiltonian. This gives
\be
S^{GF}_{\text{fixed bg}} = \int dt [\pp \dot{\phi} - \bar{H}_p]
\ee
with,
\be
\bar{H}_p = -P_a \dot{a} + H_p = \dfrac{P_a^2}{12 a} + H_p = \left( \dfrac{P_a^2}{24 a} + a^3 \Lambda \right) + \dfrac{\pp^2}{2 a^3} + a^3 V(\phi)
\ee
where it is understood that $a$ and $P_a$ are to be evaluated on the fixed background, and the only dynamical degree of freedom left is $\phi$.

On this fixed background, the term in the brackets is a fixed function of time (since $a(t),P_a(t)$ are fixed functions of $t$); the resulting physical Hamiltonian is 
\be
\label{dust-loc-ham}
\bar{H}_p = \dfrac{\pp^2}{2 a^3} + a^3 V(\phi)
\ee
which is the standard Hamiltonian of a scalar field on an FLRW background, which can be used to describe local (non-gravitational) physics. Now, since $a(t)$ is arbitrary, it is possible to choose $a(t)=$constant; this obviously gives the flat spacetime scalar field Hamiltonian. As a brief summary, what this shows is that the global dust time gauge and the choice $a(t) =constant$ gives the standard scalar field Hamiltonian.

At this stage one may ask what exactly has been achieved? If we had started from general relativity coupled to only a scalar field and set $a(t)$=constant and $p_a(t)=0$ in the canonical equations, solving the remaining equations would yield $\phi=0$ and $P_\phi=0$, i.e. the trivial solution. Thus, we see that in the dust + scalar + gravity system, choosing dust time and a background gives scalar field theory on that background. This holds for {\it any} background, not just FLRW. 

What is unique about dust time is that the physical Hamiltonian is not a square root; it provides a direct way to connect global time to local time; and produces a matter theory on a fixed background if the gravitational variables are taken to be fixed functions.

\subsection{Other clocks}

Let us look at three other choices of time that can be made in this theory, scalar field time $t=\phi$, Hubble time $t=1/H$, and York time $t= 2\pi/3\sqrt{q}= P_a/3a^2$ (where $\pi$ is the trace of the ADM momentum and $q$ is the determinant of the spatial metric). The corresponding physical Hamiltonians are 
\bea
H_p^\phi &=& \sqrt{ \dfrac{a^2 P_a^2}{12} - 2a^6 \Lambda - 2a^6 V(t) - 2 a^3 P_T },\\
H_p^H  &=& 2 \dfrac{ P_T \mp \sqrt{P_T^2 + 2 \frac{\pp^2}{t^2} \left[6 - t^2 ( \Lambda + V(\phi) ) \right] } }{\left[ 6 - t^2 ( \Lambda + V(\phi) ) \right]},\\
H_p^Y  &=& \dfrac{ P_T \pm \sqrt{P_T^2 + 2 \pp^2 \left[\frac{3}{8} t^2 - ( \Lambda + V(\phi) ) \right] } }{2 \left[ \frac{3}{8} t^2 - ( \Lambda + V(\phi) ) \right]}.
\eea

In each of these cases, it is not at all clear how to establish a correspondence with  the standard scalar field Hamiltonian.  Thus we see that it is only with certain rather special  choices of a global time gauge, volume and dust, with further simplifying assumptions, that we are able to establish a correspondence with the local Minkowski patch scalar field Hamiltonian.   

\section{Beyond Cosmology} \label{sec-beyond-cosmo}

We now ask whether we can get to a theory describing local physics starting from full general relativity coupled to matter fields without reducing to cosmology. In general, we have seen that with any arbitrary choice of global clock, it is not possible to do this, even in the restricted setting of cosmology. However, it seems possible with the dust time gauge, where the physical Hamiltonian is not a square root. We start with the full theory (Eqns. \ref{full-theory}, \ref{full-ham}) and set the gauge $T=t$. 

With this choice, the dust contribution to the diffeomorphism constraint vanishes, the dust Hamiltonian reduces to $\H_D = P_T$, and the physical Hamiltonian (density) becomes
\be
\H_p = \H_G + \H_{\phi} = \dfrac{1}{\sq} \Bigg(\pi^{ab} \pi_{ab} - \frac{1}{2} \pi^2 \Bigg) + \sq ( \Lambda - R) + \dfrac{\pp^2}{2 \sq} + \dfrac{1}{2} \sq q^{ab} \partial_a \phi \partial_b \phi + \sq V(\phi),
\ee
with the action
\be
S^{GF} = \int d^3x dt ~ [ \pi^{ab} \dot{q}_{ab} + \pp \dot{\phi} -\H_p - N^a (C^G_a + C^{\phi}_a)],
\ee
where
\be
C^{G}_a = D_b \pi^b_a, \quad C^{\phi}_a = \pp \partial_a \phi.
\ee
We note that Minkowski spacetime ($q_{ab}=\delta_{ab}, \pi^{ab} = 0 = N^a, N=1, \Lambda = 0$), along with zero scalar field ($\phi = 0 = \pp$) is a solution to the equations of motion of this theory. To make a connection with local physics, we consider a linearization of this theory about the fixed Minkowski background. The following discussion closely follows \cite{Ali:2015ftw}, where the authors consider a linearization of gravity about Minkowski spacetime in the dust time gauge (the main difference here is the addition of a scalar field). We cite their main results, show what differences occur with the addition of the scalar field, and refer to \cite{Ali:2015ftw} for technical details.

The linearization proceeds by writing,
\bea
q_{ab} &=& \delta_{ab} + h_{ab} (x,t), ~~ \pi^{ab} = 0 + p^{ab} (x,t), ~~ N^a = 0 + \xi^a (x,t), \nn \\
\phi &=& 0 + \varphi (x,t), ~~ \pp = 0 + \pv (x,t),
\eea
where $h_{ab}, p^{ab}, \xi^a, \varphi,$ and $\pv$ are small perturbations, and we expand the Hamiltonian to second order in these perturbations to obtain equations of motion at the linear order. Under this linearization, the scalar field perturbations ($\varphi, \pv$) decouple from the gravitational perturbations, and the scalar field Hamiltonian becomes,
\be
\H_{\varphi} = \dfrac{\pv^2}{2} + \dfrac{1}{2} \delta^{ab} \partial_a \varphi \partial_b \varphi + V(\varphi),
\ee
which is the usual Hamiltonian for a scalar field on a flat background. The contribution of the scalar field to the diffeomorphism constraint ($\pv \partial_a \varphi$) also vanishes at the linear order, and hence only the gravitational contribution remains.

The gravitational sector of this theory gives rise to two graviton modes, and an ultralocal scalar mode \cite{Ali:2015ftw}. Hence we find that, after fixing dust as global time in the full theory, Minkowski spacetime is a solution, and that perturbations about this fixed background Minkowski solution give rise to two graviton modes, an ultralocal scalar mode, and the usual scalar field dynamics that we would expect from local physics.

\section{Summary and Discussion} \label{sec-disc}

In the foregoing we posed and discussed the evolution of a scalar field with respect to ``global time" as defined by a canonical gauge fixing in gravity-matter system, and with respect to the unique time variable provided by the Killing vector field of Minkowski spacetime. While the resulting Hamiltonians are distinct at the structural level, we find these different times lead to very similar evolution on the time scales of terrestrial experiments. Of particular note is the dust time choice, where Poincare invariance of the scalar field remains unscathed despite the apparent intuition that any global choice of time should break it. 

We used the volume and dust time in cosmology for this comparison of scalar field dynamics. There is of course no known general method available for extracting a unique time variable---if there were one, we would have a solution to the problem of time, as for instance in the parametrized particle toy model where the Hamiltonian constraint is linear in one momentum. Dust time is apparently the closest one can come to this feature for gravity-matter systems. And dust time retains global Poincare invariance despite making an explicit time choice even in the full theory (i.e. without reduction to cosmology), a remarkable feature already noted in  \cite{Ali:2015ftw} where linearized gravity is developed in the dust time gauge. 

We also noted that ``evolving coupling constants" arise naturally in the volume time gauge. The basic ``mechanism" is straightforward: matter potentials in the Hamiltonian have the universal form $\sqrt{q}V(\phi) = a^3 V(\phi) = t V(\phi)$ in homogeneous cosmology in the volume time gauge. Hence all coupling constants in $V(\phi)$ acquire time dependence. We described the details of this mechanism, where a canonical transformation is necessary to bring the free part of the scalar Hamiltonian to its familiar form. Our main result following such an analysis is that all coupling constants are driven to zero with evolution in volume time. Thus, while one might be tempted to argue that cosmological evolution makes all matter interactions ``asymptotically free,"  we will not do so here. 

Lastly we note another feature of global time that is relevant for the so-called cosmological constant ($\Lambda$)/vacuum energy problem. Its conventional formulation is in the framework of effective field theory on Minkowski spacetime, where a cosmological constant is generated in perturbation theory; (see e.g. \cite{rugh2000,koberinski2023} for foundational discussion).  From the perspective of a global time, the corresponding physical matter-gravity Hamiltonian  contains $\Lambda$ as a coupling constant. In the volume time gauge the physical Hamiltonian is not linear in $\Lambda$, whereas in the dust time gauge it is linear \cite{Husain:2015dxa, Hassan:2019xej}. The vacuum energy on the other hand would be, by the usual definition, the ground state energy $E_0$ of such a Hamiltonian; this quantity that would be a function of all coupling constants $g_i$, including $\Lambda$, i.e. $E_0(\Lambda, g_1,g_2,\cdots g_n)$. Thus, from the perspective of global time,  there is no reason to associate vacuum energy (so defined)  with the cosmological constant. This observation  provides a point of contrast on the consequences of global time in a gravity-matter system on the one hand, and Minkowski Killing time for matter fields on flat spacetime on the other.

\noindent{\bf Acknowledgments} The work of VH was supported in part by a Discovery Grant from the Natural Sciences and Engineering Research Council of Canada. 

\bibliography{ChangingConstants2}

\end{document}